\documentclass[artical]{IEEEtran}
\usepackage{epsfig,graphics,epsf,color,amsmath,balance,cite,authblk}
\usepackage{graphicx}
\usepackage{psfrag}
\usepackage{stfloats}
\usepackage{amsmath}
\usepackage{amssymb}
\usepackage{array}
\usepackage{algorithm}
\usepackage{algorithmic}
\usepackage{caption}

\providecommand{\keywords}[1]{\textbf{\textit{Index terms---}} #1}

\begin{document}
%
% paper title
% Titles are generally capitalized except for words such as a, an, and, as,
% at, but, by, for, in, nor, of, on, or, the, to and up, which are usually
% not capitalized unless they are the first or last word of the title.
% Linebreaks \\ can be used within to get better formatting as desired.
% Do not put math or special symbols in the title.
\title{ An Incremental Redundancy HARQ Scheme for Polar Code }

% author names and affiliations
% use a multiple column layout for up to three different
% affiliations
%\author{\IEEEauthorblockN{Liang~Ma}
%\IEEEauthorblockA{Huawei Technology CO., LTD\\Shanghai Institution\\
%Shanghai, China 201206\\
%Email: maliang9@huawei.com}
%\and
%\IEEEauthorblockN{Yuejun~Wei}
%\IEEEauthorblockA{Huawei Technology CO., LTD\\Shanghai Institution\\
%Shanghai, China 201206\\
%Email: weiyuejun@huawei.com}
%\and
%\IEEEauthorblockN{Yuejun~Wei}
%\IEEEauthorblockA{Huawei Technology CO., LTD\\Shanghai Institution\\
%Shanghai, China 201206\\
%Email: weiyuejun@huawei.com}
%}

% conference papers do not typically use \thanks and this command
% is locked out in conference mode. If really needed, such as for
% the acknowledgment of grants, issue a \IEEEoverridecommandlockouts
% after \documentclass

% for over three affiliations, or if they all won't fit within the width
% of the page, use this alternative format:
%
\author{\IEEEauthorblockN{Liang~Ma\IEEEauthorrefmark{1},
Jie~Xiong\IEEEauthorrefmark{1},
Yuejun~Wei\IEEEauthorrefmark{1} and
Ming~Jiang\IEEEauthorrefmark{2}}\\
\IEEEauthorblockA{\IEEEauthorrefmark{1}Huawei Technology CO., LTD\\
Shanghai Institution,Shanghai, China 201206\\
Email: \{maliang9,liuxiaojian5,pico.xiong,zhengchen1,weiyuejun\}@huawei.com \\ }
\IEEEauthorblockA{\IEEEauthorrefmark{2}SouthEast University\\
No.9, Mozhoudong Road, Jiangning District, Nanjing, China 211111\\
Email: Jiang\_ming@seu.edu.cn \\}}

% use for special paper notices
%\IEEEspecialpapernotice{(Invited Paper)}

% make the title area
\maketitle

% As a general rule, do not put math, special symbols or citations
% in the abstract
\begin{abstract}
A polar code extension method which supports incremental redundancy hybrid ARQ(IR-HARQ) is proposed in this paper. By copying information bits to proper positions of the extend part, the extended polar code can give additional protection for the bits weakly protected by the first transmission. A comparison between the proposed algorithm with directly generated polar code, LTE turbo code, and some other IR-HARQ supporting polar code is given. Simulation results show that the proposed algorithm has nearly the same performance as directly generated polar code, which is chosen as base line for comparison.
\end{abstract}


% For peer review papers, you can put extra information on the cover
% page as needed:
% \ifCLASSOPTIONpeerreview
% \begin{center} \bfseries EDICS Category: 3-BBND \end{center}
% \fi
%
% For peerreview papers, this IEEEtran command inserts a page break and
% creates the second title. It will be ignored for other modes.
\IEEEpeerreviewmaketitle

\section{Introduction}
Polar code, proposed by Arikan \cite{Arikan09_0}\cite{Arikan09_1}, is the first class of channel code which is proven that can achieve the symmetric capacity of a binary-input discrete memoryless channel for an infinite code length under successive cancellation (SC) decoding. Later, successive cancellation list (SCL) \cite{Tal11} and successive cancellation stack (SCS) \cite{Niu12} decoders are introduced to improve the performance for a finite-length code. With CRC aided, polar code using SCL decoder outperforms previous channel codes such as turbo code or LDPC code. Polar code shows great potential for communication system in the future, and was chosen for 5G control channel recently.

Wireless system operates in time-varying channels, so it requires flexible and adaptive transmission techniques. Hybrid ARQ transmission schemes which combine the conventional ARQ with forward error correction code are very useful for scenarios where the quality of the communication channel is unknown. Incremental redundancy hybrid ARQ (IR-HARQ) achieves better performance,  so it has been adopted by a number of standards for mobile phone networks. It usually operates as follows. The user data bits are firstly encoded by a low rate code, referred to as the mother code. Then, only a part of mother code is sent and the receiver tries to decode by these selected bits. If the decoding attempt fails, additional bits will be sent. The receiver combine the bits retransmitted with those bits previously received, and decode them as a relatively lower rate code. This procedure is repeated until successfully decoding or all of the mother code is sent \cite{Andriyanova09}.

However, it is hard for polar code to design an IR-HARQ strategy by the scheme above. Due to the highly structured nature of polar code, a puncture method which can maintain a capacity-achieving performance of its punctured code is still not found. Nor is it clear how to incrementally add coding bits to a high rate polar code and make sure the final low rate code is capacity-achieving. In \cite{Niu13}-\cite{Miloslavskaya15}, kinds of information set optimizing methods based on different puncture pattern were proposed, but these methods cannot be used for IR-HARQ for which the information sets should not be changed after puncturing. In \cite{Saber15}\cite{El-Khamy15}, puncture sets were searched step by step using a greedy search algorithm which can make sure that the puncture pattern were good for current step, but the overall puncture pattern was not optimal. In \cite{Li15}\cite{Hong15}, a scheme called incremental freezing was proposed. By this scheme, a high rate polar code was sent first. If the decoding failed, the information bits sent on less reliable channels are re-encoded and sent by a low rate polar code. After the low rate code decoded, these bits were copied back to the previous code which made its rate to be lower. By this scheme, a part of coding gain can be got but not all, since retransmitted code has to be decoded separately before combining,. For IR-HARQ, like LTE turbo code, only combining soft information before decoding can be capacity-achieving.

Our purpose is to design an IR-HARQ scheme for polar code like LTE turbo or Raptor-like LDPC, which means, no matter how many bits to be decoded, receiver should decode them as an integrated low code rate code word. The combining code words should performance as good as the codes generated directly with the same code rate. In this paper, we will show how our proposed scheme can do this.

The rest of the paper is organized as follows. In Section \uppercase\expandafter{\romannumeral2}, we present some background knowledge of polar codes. In Section \uppercase\expandafter{\romannumeral3}, we present our algorithm for polar code IR-HARQ which can get full part of coding gain. Section \uppercase\expandafter{\romannumeral4} presents the simulation results for different IR-HARQ scheme. It also compares the performance of proposed scheme with LTE turbo code. Conclusion and topics for future research are provided in Section \uppercase\expandafter{\romannumeral5}

\section{Background}
\subsection{Polar code}

A polar Code is a linear block constructed recursively from many basic polarization units F, where $F = \left[ {\begin{array}{*{20}{c}}
1&0\\
1&1
\end{array}} \right]$. For a code with length $N = {2^n}$, the generator matrix ${G_N} = {F^{ \otimes n}}$ and $ \otimes $ denotes the Kronecker product.

In \cite{Arikan09_0}, it was proved that the recursively structure brought a kind of phenomenon called channel polarization. For a general polarization unit  $(W_N^{(i)},W_N^{(i)}) \to (W_{2N}^{(2i - 1)},W_{2N}^{(2i)})$, $n \ge 0,N = {2^n},1 \le i \le N$, while N denote the length and i denote the index of information bits to be encoded.

It can be proved that when N independent channels goes through the butterfly patterns, some of them become more reliable and others become unreliable. For any B-DMC channel, as N goes to infinity through powers of two, the capacity for part of the channels goes to 1 and others goes to 0. The main idea of polar code is send data only through those channels whose capacity is near 1, and send frozen bits (known bits for sender and receiver) through those channels whose capacity is near 0. Details of these symbols mentioned above can be found in \cite{Arikan09_0}.

\subsection{Channel capacity caculate}

Encode of polar code is very easy to be described by the butterfly figure. Firstly, the capacity of each bit has to be calculated. For B-DMC channel, Arikan gives a way to computed it by recursive relations:
\begin{equation}
\label{e03}
I(W_N^{(2i - 1)}) = I{(W_{N/2}^{(i)})^2}
\end{equation}
\begin{equation}
\label{e04}
I(W_N^{(2i)}) = 2I(W_{N/2}^{(i)}) - I{(W_{N/2}^{(i)})^2}
\end{equation}

For AWGN channel, Gaussian approximation is used to estimate channel capacity easily \cite{Mori09}\cite{Chung01}. Assume the input signal $a_N^{(i)}$ is a Gaussian distribution. The mean value and variance $(m_{2N}^{(2i - 1)},2m_{2N}^{(2i - 1)})$ and $(m_{2N}^{(2i)},2m_{2N}^{(2i)})$ of output signal $a_{2N}^{(2i - 1)}$ and $a_{2N}^{(2i)}$ equal to:
\begin{equation}
\label{e05}
m_{2N}^{(2i - 1)} = {\varphi ^{ - 1}}(1 - {[1 - \varphi (m_N^{(i)})]^2})
\end{equation}
\begin{equation}
\label{e06}
m_{2N}^{(2i)} = 2m_N^{(i)}
\end{equation}

Function $\varphi ()$:
\begin{equation}
\label{e07}
\varphi (x) = \left\{ {\begin{array}{*{20}{c}}
{\begin{array}{*{20}{c}}
{\sqrt {\frac{\pi }{x}} (1 - \frac{{10}}{{7x}})\exp ( - \frac{x}{4})}\\
{\exp ( - 0.4527{x^{ - 0.86}} + 0.0218)}
\end{array}}&{\begin{array}{*{20}{c}}
{(x \ge 10)}\\
{(0 < x < 10)}
\end{array}}
\end{array}} \right.
\end{equation}

Assume the information sequence is denoted as $\hat u = \{ {u_0},{u_1}, \ldots ,{u_N}\} $. Through recursively applying (\ref{e05}) and (\ref{e06}), $I(W_N^{(i)})$ for each bit ${u_i}$ can be calculated. Then, set the bits with least channel capacity to be zero, and let the other k bits to be variable. Codewords $\hat x$ can be calculated as
\begin{equation}
\label{e08}
\hat x = \hat u{G_N}
\end{equation}
${G_N}$ denotes the generator matrix for polar code.

\section{IR-HARQ for Polar code}
\subsection{Illustration of proposed algorithm}
For LTE turbo code, puncture is an easy method to achieve rate compatible. However, puncture is not a good way for polar code, because it will break the uniform structure of polar units and the channel polarization distribution will change after that. Another way to achieve rate compatible is extending, since Raptor-like LDPC used this method to construct a low rate code from high rate code. The problem for extending polar code is: how to make sure the information bits are still sent by the good channels after extending?

\begin{figure}[tb]
    \begin{center}
      \epsfxsize=7.0in\includegraphics[scale=0.55]{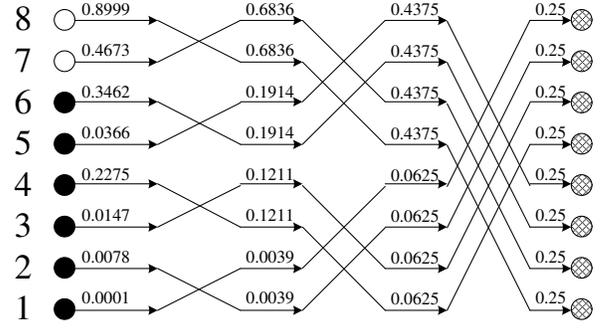}
      \caption{An Example for a (8, 6) polar code  \label{First Transmission Example}}
    \end{center}
\end{figure}

For easy understanding, we will give an example to illustrate the main idea of proposed algorithm. As Fig.\ref{First Transmission Example} shows, we send a (8, 6) polar code at first. The channel is a BEC with erasure probability p=0.25, and the channel capacity for each bits can be calculated by (\ref{e03}) and (\ref{e04}). We write these erasure probabilities on the figure. If we want to set 6 variable bits and 2 frozen bits from bit No. 1 to bit No. 8, bit No. 7 and bit No. 8 should be set to be frozen bits for they have a higher erasure probability than others. Define "capacity sequence" to be the bit sequence ordered by its channel capacity from higher to lower (For BEC, it should be the erasure probability from lower to higher). ${a_1},{a_2}, \cdots ,{a_n}$ denote the bits before encoding and ${b_1},{b_2}, \cdots ,{b_n}$ denote the bits after encoding. The capacity sequence for the (8, 6) polar code should be ${a_1},{a_2},{a_3},{a_5},{a_4},{a_6},{a_7},{a_8}$.

\begin{figure*}[tb]
    \begin{center}
      \epsfxsize=7.0in\includegraphics[scale=0.5]{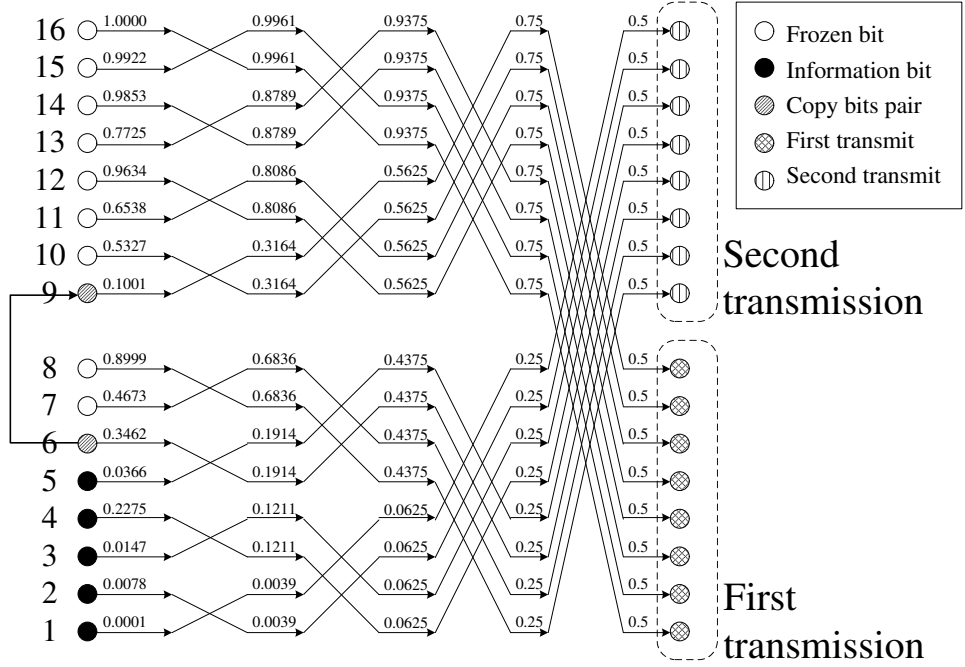}
      \caption{An Extenting Example from a (8, 6) polar code to a (16, 6) polar code   \label{An_Example of Extension}}
    \end{center}
\end{figure*}

Assuming channel erasure probability p=0.5 now and the (8, 6) polar code was decoded unsuccessfully. For IR-HARQ, the (8, 6) polar code need to be extended to a (16, 6) polar code, so the code can work for lower SNR. As Fig.\ref{An_Extenting_Example} shows, the original (8, 6) polar code is extended to a (16, 6) polar code by recursively add 8 new bits to the original code. Now the problem is, the value for bits ${a_1}$ to ${a_8}$ and ${b_1}$ to ${b_8}$ cannot be changed, but some of the information bits were not set on the good position for a (16, 6) polar code. To solve the problem, first we recalculate the erasure probabilities for the (16, 6) polar code. Then, we check the erasure probability from ${a_9}$ to ${a_16}$. It is found that if we include the extension part, ${a_9}$ got a lower erasure probability than ${a_6}$ and ${a_4}$ which were set to be information bits for the first transmission. That means, it is better to send bit by ${a_9}$ rather than ${a_6}$ or ${a_4}$. Since ${a_6}$ has a higher erasure probability than ${a_4}$, we copy the value of ${a_6}$ to ${a_9}$ and set ${a_9}$ to be an information bit for the (16, 6) polar code. The other bits from ${a_10}$ to ${a_16}$ has a higher erasure probability than the information bits for the first transmission, so we set them to be frozen bits.

After the extending, we got two part of polar code. As Fig.\ref{An_Example of Extension} shows that, one part for the first transmission and the other part for the second transmission. When retransmission occurs, the extension part should be combined with the first transmitted part and decoded as a (16, 6) polar code. When the decoding progress goes to ${a_9}$, it is treated as an information bit. Then, when the decoding progress goes to ${a_6}$, since it contains the same information as ${a_9}$, ${a_6}$ is decoded as a frozen bit and use the value of ${a_9}$ to choose its path (0 or 1).

Here, we define the bits pair like ${a_6}$ and ${a_9}$ as "copy bits pair". For polar IR-HARQ, we need to find these "copy bits pair" when extent a N length polar code into a 2N length polar code. Here, we will describe our method in AWGN channel which is similar to the example shows above. ${a_i}$ denotes the bits before polar encoding and ${b_i}$ denotes the bits after polar encoding. ${C_{N,{a_i}}}$ donates the channel capacity for ${a_i}$ calculated by N length polar code. ${C_{2N,{a_i}}}$ donates the channel capacity for ${a_i}$ calculated by 2N length polar code.

Step 1, we use Gaussian approximation to calculate capacity sequence for the N length polar code as \[{{\bf{a}}_N} = \{ {a_{{i_0}}},{a_{{i_1}}}, \cdots ,{a_{{i_{N - 1}}}}\} ,0 < {i_k} \le N\]

If total number of information bits are k, top k bits in ${{\bf{a}}_N}$ sequence are set as information bits.

Step 2, we calculate the capacity sequence for the 2N length polar code. The capacity sequence of the bits belonging to the first transmission part is \[{{\bf{a}}_{2N}} = \{ {a_{{p_0}}},{a_{{p_1}}}, \cdots ,{a_{{p_{N - 1}}}}\} ,0 < {p_k} \le N\]

The capacity sequence of the the bits belonging to the retransmission part is \[{{\bf{r}}_{2N}} = \{ {r_{{q_0}}},{r_{{q_1}}}, \cdots ,{r_{{q_{N - 1}}}}\} ,N + 1 < {q_k} \le 2N\]

${C_{2N,{a_i}}}$ and ${C_{2N,{r_q}}}$ donates the channel capacity for ${a_i}$ and ${r_q}$ calculated by 2N length polar code.

Step 3, search ${{\bf{a}}_{2N}}$ and ${{\bf{r}}_{2N}}$. Find the smallest ${C_{2N,{a_i}}}$, whose corresponding ${a_i}$ belongs to an information bit for the first transmission. Then, find the smallest ${C_{2N,{r_q}}}$ which is bigger than ${C_{2N,{a_i}}}$. Set ${a_i}$ and ${r_q}$ as a copy bits pair and remove ${a_i}$ and ${r_q}$ from ${{\bf{a}}_{2N}}$ and ${{\bf{r}}_{2N}}$.

Step 4, repeat step 3, until no ${C_{2N,{r_q}}}$ in ${{\bf{r}}_{2N}}$ satisfies the criterion in step 3.

By the four steps, all copy bits pair are found. For each copy bits pair, one bit locates in the first transmission part, and the other bit locates in the retransmission part. We copy the value of the bits in the first transmission part to its corresponding partner. For the first transmission, the bit of a pair in the first transmission part can be decoded as an information bit. For the retransmission, the bit of a pair in the retransmission should be decoded as an information bit, while its partner can be decoded as a frozen bit and using former decoded result for choosing path(0 or 1).

We give the operation in a pseudo code form.
\noindent
\rule[0pt]{9cm}{0.1em}
\begin{algorithmic}
\STATE Set code length to N
\FOR{i in N}
\STATE Calculate the channel capacity for the i-th bit as ${C_{N,{a_i}}}$
\ENDFOR

\STATE Set code length to 2N
\FOR{i in 2N}
\IF{$i \le N$}
\STATE Calculate the channel capacity for the i-th bit as ${C_{2N,{a_i}}}$
\ELSE
\STATE Calculate the channel capacity for the i-th bit as ${C_{2N,{r_i}}}$
\ENDIF
\ENDFOR

\FOR{all ${C_{2N,{a_i}}}$ in increasing order}
\IF{$a_i$ is an information bit for 1st transmission}
\FOR{all ${C_{2N,{r_j}}}$}
\IF{exist ${C_{2N,{r_j}}}$, ${C_{2N,{r_j}}} > {C_{2N,{a_i}}}$ }
\STATE Find smallest ${C_{2N,{r_k}}}$, ${C_{2N,{r_k}}} > {C_{2N,{a_i}}}$
\STATE Set ${C_{2N,{r_k}}}$ and ${C_{2N,{a_i}}}$ as a copy pair
\ELSE
\STATE Break
\ENDIF
\ENDFOR
\ENDIF
\ENDFOR
\end{algorithmic}
\noindent
\rule[0pt]{9cm}{0.1em}

\subsection{Method for multiple retransmission}
\begin{figure}[tb]
    \begin{center}
      \epsfxsize=7.0in\includegraphics[scale=0.42]{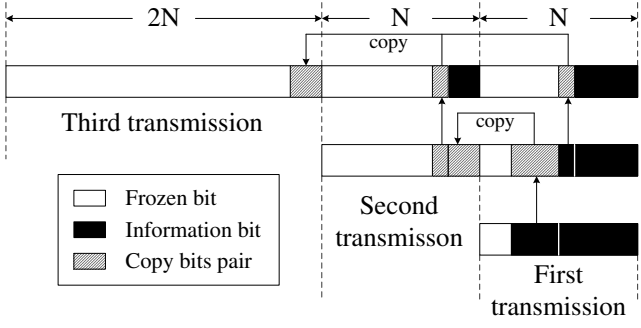}
      \caption{IR-HARQ for duplicate retransmissions  \label{duplicate_retransmissions}}
    \end{center}
\end{figure}
Part A has illustrated the main idea for one time retransmission, and in this part the method for multiple retransmission will be given. Fig.\ref{duplicate_retransmissions} is an example of the proposed IR-HARQ scheme. Assume there is a polar code IR-HARQ system which sends a (N, k) polar code at the first transmission and the system send N and 2N redundant bits at the second and the third time. (If there are more repeat requests, the retransmitted bits grow according to N multiply power of 2 like 4N, 8N, 16N,...)

Assume we have an N length polar code, and the code is extended to a 2N length polar code for 2nd transmission. If the 3rd transmission are needed, we can construct a 4N polar as the method shown in part A. First, the information bits for the 2N polar should be found. For the bits belonging to information bits for the first transmission and not belonging to a copy bits pair, the bits are treated as information bits. For the bits belonging to a copy bits pair and not belonging to the first transmission part, the bits are treated as information bits. The other bits belonging to the the 2N polar code are treated as frozen bits.

After all information bits for the 2N polar code are found, we can use similar method as section A to find all copy bits pair for 4N polar code. For each pair, duplicate the value of the bits belonging to the 2N polar code to its partner. The 4N polar code can be got.

Decoding scheme is also similar as section A. For each copy bit pair, both bits contain the same value. When decode sequence goes to a copy bit pair, check the value of its partner. If its partner is not been decoded yet, the bit is treated as an information bit. If its partner is already decoded, the bit is handled as a frozen and uses its partner's value to choose path.

\section{Simulation Results}
In this section we present Monte-Carlo simulation results to evaluate the performance of different IR-HARQ schemes. We also give a comparison between proposed scheme and LTE turbo code.

\begin{figure}[tb]
    \begin{center}
      \epsfxsize=7.0in\includegraphics[scale=0.4]{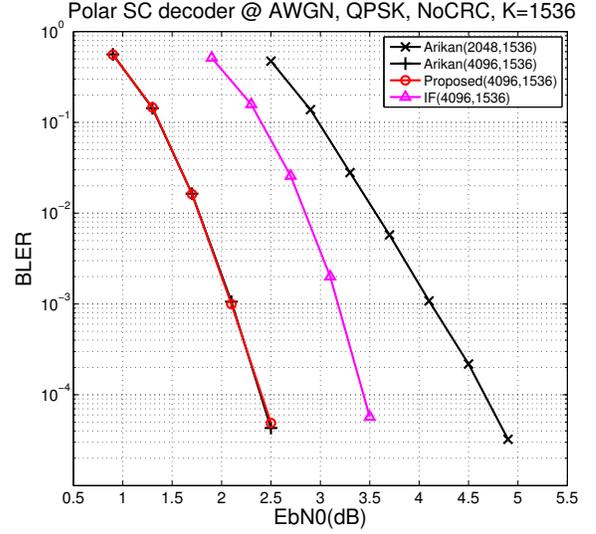}
      \caption{Performance of proposed algorithm with k=1536  \label{dPolarSCdecoderHARQ(K=1536)}}
    \end{center}
\end{figure}

We assume code is transmitted on AWGN channel by QPSK and decoded by SC decoding. The capacity sequence is calculated by Gaussian approximation. The estimate SNR is set to 4 dB for the first transmission and 1 dB for the second transmission.
Fig.\ref{dPolarSCdecoderHARQ(K=1536)} shows the block error rate (BLER) performance of polar codes when k=1536. For comparison, simulation curve of incremental freezing (IF) and directly generated Arikan polar code is given at the same time. Compared with directly generated polar code when k=1536, the proposed algorithm performs as well as it, while the incremental freezing algorithm is 1.1 dB worse.

\begin{figure}[tb]
    \begin{center}
      \epsfxsize=7.0in\includegraphics[scale=0.4]{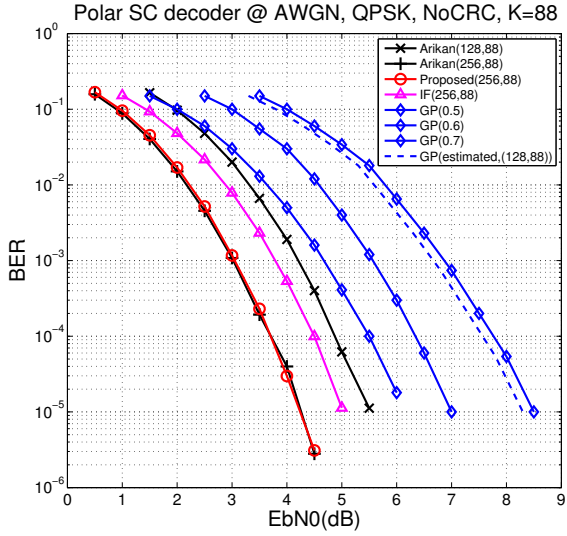}
      \caption{The BER comparision between different HARQ algorithm with k=88  \label{dPolarSCdecoderHARQ(K=88)}}
    \end{center}
\end{figure}

Fig.\ref{dPolarSCdecoderHARQ(K=88)} shows the bit error rate (BER) performance of polar code when k=88. The greedy puncture (GP) algorithm is added for comparison. In this case, proposed method performs nearly the same with directly generated Arikan polar code, while incremental freezing method is 0.6 dB worse. The performance for greedy puncture deteriorates with the increase of puncture bits. When the rate goes up to 0.685 which nearly equal to (88, 256) polar code, the performance of greedy puncture is 2.3 dB worse than directly generated Arikan polar code.

\begin{figure}[tb]
    \begin{center}
      \epsfxsize=7.0in\includegraphics[scale=0.4]{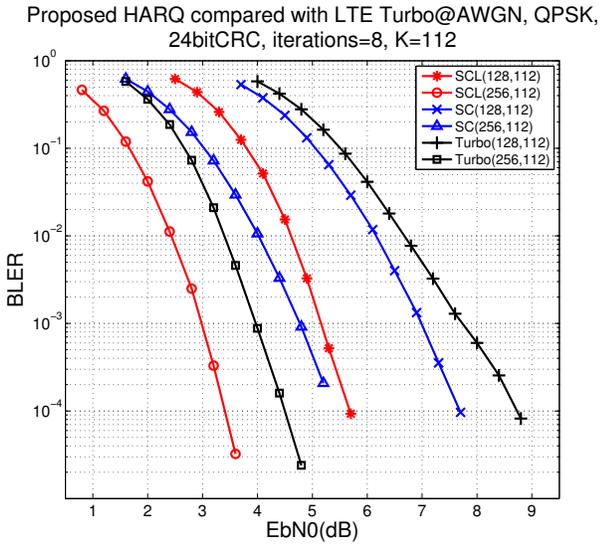}
      \caption{Comparision between proposed algorithm and LTE Turbo code  \label{PolarvsTurbo}}
    \end{center}
\end{figure}

Fig.\ref{PolarvsTurbo} shows a BLER performance comparison between the proposed method with LTE turbo code. A Max-Log-Map algorithm is applied in turbo decoding and the iteration times equal to 8. The result shows that polar code decoded by SCL algorithm outperforms LTE turbo before and after the retransmission. Polar code decoded by SC algorithm outperforms LTE turbo code before retransmission but performs worse than LTE turbo code after retransmission. This simulation result shows that by proposed algorithm, polar code using SC decoding algorithm can performs as well as turbo code in a system need IR-HARQ. If SCL decoding algorithm is used, polar code even has a better performance than turbo code.

However, there are still some disadvantages of polar code for IR-HARQ, since the proposed algorithm requires the length of retransmitting blocks grows according to multiples of 2. Compared with turbo code, it is inconvenient for polar to do a flexible rate compatible IR-HARQ. If the length for each transmitting blocks can be fix to N, we can use the idea of incremental freezing to solve this problem. We decode the retransmitting block first and combine the decoded soft value with its corresponding copy bits. When the total transmitted blocks satisfy $N \cdot {2^k}$ and k is an integer, retransmitted blocks will be combined with previous blocks and decoded as a complete low rate polar code. By this solution, full of coding gain will be got when blocks are combined and for other cases, there will be a performance loss. Fortunately, in a real wireless communication system, the average retransmitting time is very close to 1, so for most cases the gain from one time retransmission is enough for decoding successfully. Only one block which has a same length as pervious block is needed for retransmitting that is very suitable for our algorithm.

\begin{figure}[tb]
    \begin{center}
      \epsfxsize=7.0in\includegraphics[scale=0.4]{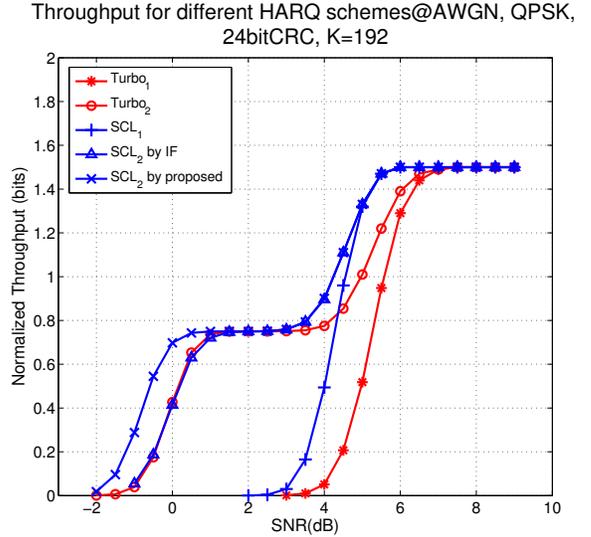}
      \caption{Throughput for different HARQ schemes (k=192)  \label{ThroughputComparision(k=192)}}
    \end{center}
\end{figure}

Normalized throughput is defined as:
\begin{equation}
\label{e14}
\Gamma  = R \cdot {\log _2}(M) \cdot (1 - BLER)/\bar t
\end{equation}
$\bar t$ represents the average number of transmissions required for successful decoding at the tested SNR. Assume max transmitting time equals to 2, Fig.\ref{ThroughputComparision(k=192)} illustrates the normalized throughput with IF and proposed schemes for polar code compared with LTE turbo code. Here, method for IF mentioned in \cite{Hong15} is used for a easier realizing. In Fig.\ref{ThroughputComparision(k=192)}, when k equals to 192 and retransmission is not used, polar code performs nearly 1 dB better than LTE turbo code. When max transmission equals to 2, it is observed that the proposed HARQ scheme provides nearly 0.7 dB gain over LTE turbo code.

\section{Conclusion}
In this paper, we introduce an IR-HARQ scheme for polar code. To improve the performance for retransmission, the proposed algorithm extend original code into a longer polar code like Raptor-like LDPC. The main idea of proposed algorithm is to find copy bits pairs and make the pairs contain same information. By this way, both original code and extended code are constructed based on calculated channel capacity which ensure the performance for both first transmission and retransmission. Simulation results show that the extended code constructed by our scheme has nearly the same performance as a directly generated polar code.

For future work, increasing the flexibility of the scheme should be focused on. For now, if the length of first transmission is N, the length of the retransmissions has to be a multiple of N, while it is almost no limit for LTE turbo code. And only when the total length of transmitted blocks satisfying $N \cdot {2^k}$ and k is an integer, the codes will be combined together and decoded as a low rate polar code. If the communication system needs multiple retransmission, the proposed algorithm need to be improved.

% conference papers do not normally have an appendix

% use section* for acknowledgment
%\section*{Acknowledgment}
%
%
%The authors would like to thank...

% trigger a \newpage just before the given reference
% number - used to balance the columns on the last page
% adjust value as needed - may need to be readjusted if
% the document is modified later
%\IEEEtriggeratref{8}
% The "triggered" command can be changed if desired:
%\IEEEtriggercmd{\enlargethispage{-5in}}

% references section

% can use a bibliography generated by BibTeX as a .bbl file
% BibTeX documentation can be easily obtained at:
% http://mirror.ctan.org/biblio/bibtex/contrib/doc/
% The IEEEtran BibTeX style support page is at:
% http://www.michaelshell.org/tex/ieeetran/bibtex/
%\bibliographystyle{IEEEtran}
% argument is your BibTeX string definitions and bibliography database(s)
%\bibliography{IEEEabrv,../bib/paper}
%
% <OR> manually copy in the resultant .bbl file
% set second argument of \begin to the number of references
% (used to reserve space for the reference number labels box)

% that's all folks
\end{document}